\documentclass[aps,twocolumn,showpacs,preprintnumbers,amssymb]{revtex4}

\newcommand{\be}{\begin{eqnarray}}
\newcommand{\ee}{\end{eqnarray}}

\newcommand{\eins}{\mbox{$1 \hspace{-1.0mm}  {\bf l}$}}
\def\bea{\begin{eqnarray}}
\def\eea{\end{eqnarray}}
\def\C{\hbox{$\mit I$\kern-.7em$\mit C$}}
\def\N{\hbox{$\mit I$\kern-.3em$\mit N$}}

\usepackage{dcolumn} 
\usepackage{bm} 

\usepackage{epsfig}

\begin{document}

\title{Stability of macroscopic entanglement under decoherence}

\author{W. D\"{u}r  and H.-J. Briegel}

\affiliation{
Sektion Physik, Ludwig-Maximilians-Universit\"at M\"unchen, Theresienstr.\ 37, D-80333 M\"unchen, Germany.}

\date{\today}

\begin{abstract}
We investigate the lifetime of macroscopic entanglement under the influence of decoherence. For GHZ--type superposition states we find that the lifetime decreases with the size of the system (i.e. the number of independent degrees of freedom) and the effective number of subsystems that remain entangled decreases with time. For a class of other states (e.g. cluster states), however, we show that the lifetime of entanglement is independent of the size of the system. 
\end{abstract}


\maketitle



The question whether or not entanglement -- a genuine feature
of quantum mechanics -- can persist in a macroscopic (i.e. ``classical")
world, has entertained quantum physicists since Erwin Schr\"odinger introduced
his notorious gedanken experiment known as ``Schr\"odinger's cat'' \cite{Sr35,Zu03}.
While entangled states of microscopic matter, such as a few atoms or ions in
a trap, can nowadays be prepared in the laboratory \cite{Fo00},
it is often argued that for a large number $N$ of particles this task
would become exceedingly difficult, since the effective decoherence rate would
grow linearly with the size of the system $N$.

In its simplest version, the argument is based on the observed
evolution of superposition states of the form 
$|GHZ\rangle \equiv 1/\sqrt{2}(|0\rangle^{\otimes N}+|1\rangle^{\otimes N})$, also called Greenberger--Horne--Zeilinger (GHZ) states, 
of a system of $N$ spins or qubits interacting with uncontrollable degrees of freedom of the environment, described e.g. by a heatbath. The rate at which this state decoheres scales indeed as $\kappa N$, where $\kappa$ is the decoherence rate of a single qubit. While this observation is correct, it is not clear whether the scaling of the decoherence rate with the size of the system $N$ is a special property of GHZ states or a general feature of all multiparticle entangled states. This means that the conclusions that are usually drawn from this observation, namely that macroscopic entanglement, i.e. entanglement between a macroscopic number of particles, necessarily becomes exponentially fragile with $N$, are questionable and will indeed be refuted in this paper. 

To this aim, we investigate the effect of decoherence on the entanglement properties of a class of multiparticle entangled states.
We consider the lifetime of entanglement between a variable number of subsystems of the system. Specifically, we consider both the time after which the distillable entanglement of the state vanishes and the time after which the state becomes separable. These lifetimes are, in general, finite. We determine the scaling behavior of these lifetimes with $N$. For GHZ states we find that an increasing decoherence rate indeed results in a lifetime of distillable $N$-party entanglement \cite{Du00,Th02} that decreases with $N$. For many other states (e.g. cluster states \cite{Ra01}), in contrast, we show that the lifetime of genuine multiparty entanglement is {\em independent} of the size of the system $N$. In particular, we find that the lifetime of any state that belongs to the class of graph states $|\phi\rangle_G$ \cite{Ra01} --which contains the GHZ and cluster state as particular cases-- is bounded from below by a quantity which depends on the maximum degree of the associated ``interaction'' graph $G$, but is independent of $N$. This implies that genuine multi-particle entanglement of a macroscopic number of particles is possible and can persist for timescales that are independent of the size of the system.

When describing multi-particle entanglement, a central notion is
the {\em partitioning} of a system of $N$ particles into $M\le N$
groups. Each group may consist of several particles, which are then
considered as a single 
subsystem with a higher-dimensional
state space. If we associate a specific spatial distribution with the
particles, as in the case of spins on a lattice, partitionings may be
chosen that correspond to a re-scaling of the size of the subsystems,
as it is used in statistical physics. In the present context, we will
be interested in the behavior of the distillable entanglement under a
coarsening of the partitions, i.e. under a re-scaling of the size of
the subsystem that is under local control of a single party. We will consider the case where the entire lattice of $N$ qubits is in a graph state and compare in particular the two cases of GHZ states and cluster states. 
On the one hand, this approach allows us to determine the effective size of the system, i.e. the number of subsystems which are still entangled after a certain time. On the other hand, we can investigate the behavior of entanglement under re-scaling in the asymptotic limit $N\to\infty$. We find that
the lifetime of the distillable entanglement of the cluster state is
largely independent of the size of the partitions, and thus the same on
all scales. For the GHZ state, in contrast, we find that the distillable entanglement
vanishes after and arbitrary short time on all scales, as long as we consider partitions of finite size.
However, if we allow the sizes of the partitions to become macroscopic
themselves (in the sense that the $N$ qubits are divided into a fixed
number of $M$ cells whose size $N/M$ grows to infinity as $N \to \infty$)
then the lifetime of this $M$-party distillable entanglement (between $M$
cells of macroscopic size) becomes finite and scales to leading order as
$1/(\kappa M)$.

Throughout this paper, we will mainly use a decoherence model corresponding to individual coupling of particles to a thermal bath in the large $T$ limit, described (in quantum information language) by individual depolarizing channels. Although we are aware of the restricted applicability of such a decoherence model, we emphasize that {\rm any} model with independent couplings of particles to the environment (which appears naturally in the case where the $N$ particles are spatially separated and thus interact with independent environments) can be mapped (by introducing additional noise) to such a depolarizing channel. This model ---in contrast to the widely used dephasing channel--- is basis independent, i.e. invariant under local unitary operations. 
In addition, in many cases it turns out that the scaling behavior of the lifetime of entanglement with the size of the system is independent of the specific  decoherence model. 

We consider depolarizing channels with noise parameter $p\equiv p(t)\equiv e^{-\kappa t}$, where $\kappa$ is a decay constant determined by the strength of the coupling to the environment and $t$ is the interaction time. The channel acting on particle $k$ is described by the completely positive map (CPM) 
\be
{\cal E}_k\rho= p(t) \rho +\frac{1-p(t)}{4} \sum_{j=0}^3 \sigma_j^{(k)} \rho \sigma_j^{(k)}, \label{whitenoise}
\ee
where $\sigma_j$ are the Pauli matrices with $\sigma_0\equiv \eins$. 
We are interested in the entanglement of a given initially pure state $|\Psi\rangle$ as a function of time. That is, the initial state suffers from decoherence described by this model and is given after a certain time $t$ by
\be
\rho(t) = {\cal E}_1{\cal E}_2 \ldots {\cal E}_N |\Psi\rangle\langle \Psi|.\label{dec}
\ee

We will use as a criterion the {\em distillable entanglement} of the system, which tells one whether it is possible to create true (irreducible) multiparticle entangled pure states. That is, we consider $N$ distinct parties each holding a particle belonging to the $N$--particle state $\rho$. The state $\rho$ is called $N$--party distillable entangled if there exists a local protocol (i.e. each of the parties acts independently on their systems) such that one can obtain from a sufficiently large number of copies of $\rho$ some true $N$--particle entangled pure state \cite{Th02}. Note that in this definition it is not necessary to specify the kind of multiparticle entangled pure state which is created, as all true $N$--party entangled pure states can be obtained from each other if several copies of the state are available \cite{Th02}. We remark that a necessary condition for $N$--party distillability is that the partial transpositions with respect to any group of parties are non--positive \cite{Du00}.

We start out by investigating the entanglement properties of GHZ states under this decoherence model. One readily finds \cite{Si02} that $|GHZ\rangle$ evolves to a state $\rho(t)$ diagonal in the GHZ--basis $\{|\Psi_{k_1k_2\ldots k_{N-1}}^\pm \rangle =1/\sqrt{2}(|k_1k_2\ldots k_{N-1} 0\rangle \pm |\bar k_1\bar k_2 \ldots \bar k_{N-1} 1\rangle)\}$, $k_j\in\{0,1\}$, with coefficients $\lambda_k^\pm$ that depend on $k\equiv\sum_j k_j$. For $k\not=0$ we have $\lambda_k^+=\lambda_k^- \equiv \lambda_k$, where   
\be
\lambda_k=[(1+p)^k(1-p)^{N-k}+(1+p)^{N-k}(1-p)^k]/2^{N+1},
\ee
while $\lambda_0^+=\lambda_0+p^N/2$, $\lambda_0^-=\lambda_0-p^N/2$. Note that $\lambda_1 \geq \lambda_2 \geq \ldots \geq \lambda_{[N/2]}$. 
The partial transposition with respect to a group $B_k$ which contains exactly $k$ parties is positive, $\rho(t)^{T_{B_{k}}} \geq 0$, if and only if \cite{Du00} 
\be 
p^N \leq 2 \lambda_{k}.\label{cond}
\ee 
For $k=1$ one observes that the threshold value $p_{\rm crit}\equiv e^{-\kappa t_{\rm crit}}$ where the partial transposition with respect to one particle becomes positive increases with $N$. This implies that for $t\geq t_{\rm crit}\equiv \tau$ the state is no longer $N$--party distillable entangled and thus the lifetime $\tau$ of true $N$--party entanglement decreases with the size of the system as expected (see Fig. 1).

\begin{figure}[ht]
\begin{picture}(230,100)
\put(-5,-5){\epsfxsize=230pt\epsffile[18 157 1107 599 ]{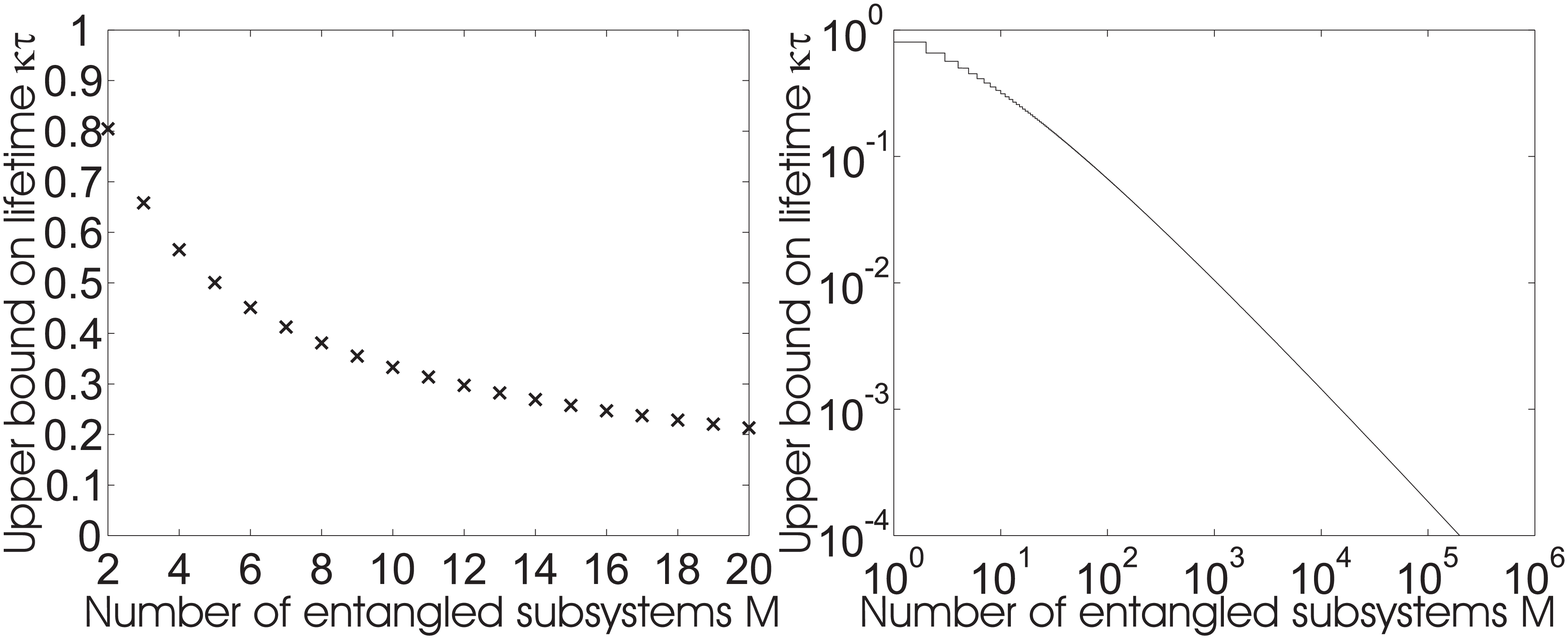}}
\end{picture}
\caption[]{Fig. 1(a): Upper bound on lifetime $\kappa \tau$ of $M$--party entanglement in systems with $N \rightarrow \infty$ particles for different $M$. Fig. 1(b): Same as Fig. 1(a) but with double--logarithmic axis. Note that the same figures are obtained for the lifetime of $N$--particle entanglement in $N$--particle systems (see text).}
\end{figure}


 We now turn our attention to a class of multiparticle entangled states, the so called graph states \cite{Ra01}. This class includes a variety of entangled states, e.g. GHZ states, cluster states and codewords of error correction codes \cite{Sc01}. 
Graph states, $|\Psi_{\mu_1\mu_2\ldots \mu_N}\rangle$, are 
defined as 
joint eigenstates of a set of $N$ commuting correlation operators $K_j$ associated with a graph $G=(V,E)$ which is a set of $N$ vertices $V$ connected in a specific way by edges $E$. The edges specify the neighborhood relation between vertices, and $K_j$ are given by
$K_j= \sigma_x^{(j)} \prod_{\{k,j\} \in E} \sigma_z^{(k)}$. 
The graph states $|\Psi_{\mu_1\mu_2\ldots \mu_N}\rangle$ fulfill the set of eigenvalue equations $K_j |\Psi_{\mu_1\mu_2\ldots \mu_N}\rangle = (-1)^{\mu_j}|\Psi_{\mu_1\mu_2\ldots \mu_N}\rangle ~\forall j$, $\mu_j \in \{0,1\}$ and form a basis in ${\cal H}=(\C^2)^{\otimes N}$. 

In the following, we consider a linear cluster state of $N$ qubits \cite{Ra01} specified by the graph $G$ with edges $(k,k+1) \forall k$ and investigate the entanglement properties of this state under the decoherence model described by Eq. (\ref{dec}). We will show that the lifetime of distillable $N$--particle entanglement is for this state {\em independent} of the size of the system. In particular, we establish a lower bound $t_{<}$ on the lifetime $t_{\rm crit}$ of distillable $N$--party entanglement for such states which is independent of the number of particles $N$. 
This is in sharp contrast to the behavior of GHZ states.     

In order to distill $N$--party entangled states, it is sufficient that maximally entangled pairs between all neighboring parties can be created, which we will show in the following. We emphasize that we use the distillability of neighboring pairs only as a tool to prove $N$--party distillability and it does not mean that the entanglement of the cluster state were in some sense only ``bipartite''. We make use of the following properties of graph states: (i) Measuring all but two neighboring particles, say $k,j$ of a graph state $|\Psi_{\vec 0}\rangle$ in the eigenbasis of $\sigma_z$ results in the creation of another graph state with only a single edge $\{k,j\}$ \cite{He03}. The resulting state of particles $k,j$ is up to local $\sigma_z$ operations equivalent to a maximally entangled state of the form $|\Phi\rangle \equiv 1/\sqrt{2}(|0\rangle_x|0\rangle_z + |1\rangle_x|1\rangle_z)$ , where $|k\rangle_x$ [$|k\rangle_z$] denote eigenstates of $\sigma_x$ [$\sigma_z$] respectively. (ii) The action of a depolarizing channel ${\cal E}_k$ on a graph state can equivalently be described by a map ${\cal M}_k$ whose Kraus operators only contain products of Pauli matrices $\sigma_z$ and the identity, where here $\sigma_z$ may act on particles $k$ and its neighbors, i.e. particles which are (in the corresponding graph) connected by edges to particle $k$. This follows from the fact that $\sigma_x^{(j)}|\Psi_{\vec \mu}\rangle = (-1)^{\mu_j}\sigma_x^{(j)} K_j |\Psi_{\vec \mu}\rangle$, where $\sigma_x^{(j)} K_j$ is an operator which contains only products of $\sigma_z$ operators at neighboring particles of particle $j$, and the identity otherwise. Similarly, the action of $\sigma_y^{(j)}$ on graph states is up to a phase factor equivalent to the action of an operator which contains only products of $\sigma_z$ operators acting on particle $j$ and all its neighbors. 

A sufficient condition when bipartite entanglement between two neighboring particles, say particle $k$ and $k+1$, can be created from $\rho(t)$ can be found straightforwardly. To this aim, one performs measurements in the eigenbasis of $\sigma_z$ on all but particles $k$ and $k+1$ (We remark that measurements on all neighboring particles of particles $k,k+1$ would also be sufficient). From properties (i) and (ii) follows that these measurements commute with the action of the CPM on the cluster state. The resulting state can thus be described by ${\cal M}_1{\cal M}_2\ldots{\cal M}_N |\tilde\Phi\rangle_{k,k+1}\langle\tilde\Phi| \otimes |\chi\rangle\langle\chi|$, where $|\chi\rangle$ is a state of the remaining $(N-2)$ particles, and  $|\tilde\Phi\rangle$ is a maximally entangled state equivalent up to $\sigma_z$ operations to $|\Phi\rangle$. It is important to note that ${\cal M}_j$ acts trivially on particles $k$ and $k+1$ if $j\notin \{k-1,k,k+1,k+2\}$ since any kind of errors in graph states only effect the corresponding particles and/or its neighbors (see (ii)). This implies that the resulting state $\rho_{k,k+1}$ of particles $k$ and $k+1$ after tracing out the remaining particles can be obtained by considering only the (reduced) action of ${\cal M}_{k-1},{\cal M}_{k},{\cal M}_{k+1},{\cal M}_{k+2}$ on the state $|\tilde\Phi\rangle\langle\tilde\Phi|$. One has that $\rho_{k,k+1}$ is distillable if its partial transpose is non--positive \cite{Ho97} and finds a threshold value $p_{<}=0.717$ ($\kappa t_{<}=0.3327$). We emphasize that this threshold value is {\em independent} of $N$, as only errors acting on the direct neighborhood of the pair of particles in question influences the threshold value. Since this method allows one to distill maximally entangled pairs between arbitrary pairs of neighboring particles, we have that for $p\geq p_{<}$ the linear cluster state remains $N$--party distillable, independent of the number of particles $N$.

We point out that above results are not restricted to linear cluster states but similarly hold for all graph states associated with some lattice geometry. To be specific, if we consider a family of graph states whose maximum degree does not increase with the number of particles $N$ and two neighboring vertices $k$ [$j$] of the corresponding graph with disjoint sets of $n_k+1$ [$n_j+1$] neighbors respectively, one can distill a maximally entangled pair between the two parties if $p^2/4[(1+p^{n_k})(1+p^{n_j})]+(1-p^2)/4 > 1/2$. A similar expression can be obtained for more general graphs where $k$ and $l$ may have common neighbors. 
It turns out that if $m=d_k+d_j-2$, where $d_i$ is the degree of vertex $i$, then for $\kappa t \leq -\ln(\frac{1}{2})/([\frac{m}{2}]+2)$ the state is certainly distillable.
For cluster states corresponding to a regular 2D [3D] lattice one finds $p_{<}=0.8281$ ($\kappa t_{<}=0.1886$) [$p_{<}=0.8765$ ($\kappa t_{<}=0.1318$)] respectively.    
This dependence on the degree is consistent with the results obtained for GHZ states, which correspond to a graph with edges $(1,k) \forall k$. The degree of this graph increases with $N$ and thus the corresponding lifetime decreases. However, we emphasize that the dependence on the degree of the graph only applies to the lower bound $t_{<}$ established with this specific purification method and the general dependence of the lifetime of an arbitrary graph state on the (degree of the) graph is presently unknown. 

We remark that the observed behavior, i.e. that the lifetime of multiparticle entanglement for cluster (and similar) states is independent of the size of the system, also holds for more general decoherence models. This follows from the fact that ---similar to (ii)--- the action of {\em any} CPM acting on graph states describing an arbitrary decoherence process can be estimated by a CPM whose Kraus operators only contain products of $\sigma_z$ operators and the identity and thus the measurement (i) still commutes with the CPM. To this aim, we apply after the application of the CPM a local depolarization procedure which maps arbitrary density operators to operators diagonal in the graph state basis without changing the diagonal elements \cite{Du03}. When restricted on graph states, the resulting action of the initial CPM (given by $\sum_{k,l} a_{k,l} O_k \rho O_l$, where $O_k,O_l$ are products of Pauli operators)) can then be described by a CPM specified by $\sum_k a_{k,k} O_k \rho O_k^\dagger$, where all operators $O_k$ can be expressed in terms of products of $\sigma_z$ operators. 
 Only operators $O_k$ which act non--trivially on particles $k,k+1$ or their neighbors in the graph affect the resulting maximally entangled pair after the measurement (i), leading again to a threshold value which is independent on the size of the system for all those decoherence models where the number of such operators $O_k$ is independent of $N$. This is for instance the case if each $O_k$ acts non--trivially on a finite, localized number of subsystems.


In the following we investigate the effective size of the system that remains entangled during the decoherence process. The fact that $N$--party entanglement vanishes after a certain time does not imply that all entanglement has disappeared. We consider partitions of the $N$--particle state into $M$ groups, where now parties within one group are allowed to perform joint operations. For states associated to some lattice geometry, specific partitions correspond to a re-scaling of the size of the subsystems. A $N$--particle state is called $M$--party distillable entangled if there exists some partitioning into $M$ parties ($M$--partitioning) such that the state is distillable to some true $M$--party entangled pure state \cite{Du00}. If the state is not $N$--party distillable entangled, it can well be that it is $M$--party distillable entangled for some $M<N$. We will be interested in the lifetime of $M$--party distillable entanglement. 

For GHZ states, one can indeed determine analytically the lifetime of $M$--party entanglement for any $M$. Recall the condition for positivity of partial transposition with respect to a group of $k$ parties given by Eq. (\ref{cond}). Since $\lambda_j \geq \lambda_i$ for $j\leq i$, we have that the group containing the fewest number of parties determines the threshold value for which the state is no longer $M$--party distillable, as the corresponding partial transposition with respect to this group is the first one to become positive. Thus $M$--party entanglement corresponding to a specific $M$--partitioning has longest lifetime if all groups have (approximately) the same size. For a minimal group size of $m$ particles, a $N$--particle GHZ state can contain at most $M\equiv [N/m]$ such groups of size $m$. This allows one to obtain the maximum lifetime of $M$--party entanglement which is determined by Eq. (\ref{cond}) with $k=m$.

One obtains an upper bound on the lifetime of $M$--party entanglement if one approximates $\lambda_m$ by some $\tilde\lambda_m < \lambda_m$ and chooses $\tilde\lambda_m \equiv (1-p)^m(1+p)^{N-m}/2^{N+1}$. In this case, the partial transposition with respect to a group of $m$ parties is certainly positive if $2\tilde\lambda_m \geq p^N$, which can be rewritten as $m \leq N \log[2p/(1+p)]/\log[(1-p)/(1+p)]$. Using that $M= [N/m]$ , we have that a $N$--particle state is certainly no longer $M$--party entangled if 
\be
 M \geq [\log(1-p)-\log(1+p)]/[\log(2p)-\log(1+p)].\label{condM}
\label{thresholdGHZ1}
\ee
On the one hand, Eq. (\ref{condM}) ---illustrated in Fig. 1--- provides an upper bound on the lifetime $\kappa \tau_M$ of $M$--party entanglement in the system. On the other hand, for a fixed time $t$ Eq. (\ref{condM}) allows one to determine the maximum $M$ of distillable multiparty entanglement remaining in the system. One observes (see Fig. 1) that $M$ rapidly decreases with $t$. For $\kappa t \ll 1$, one finds that $M \approx -2\log(\kappa t)/(\kappa t)$, while for $\kappa t> 0.8049$, we have that also $2$--party entanglement disappears and the state becomes fully separable as all partial transposes are positive (which is a sufficient condition for separability for states diagonal in the GHZ state basis \cite{Du00}). 
We also point out that this bound on the maximal size of $M$--party entanglement is independent on the number of particles $N$. That is, even if one considers groups of size $m \rightarrow \infty$ (for a total number of particles $N \rightarrow \infty$), the maximum number $M$ of such groups which remain entangled after a certain time $t$ is finite.  While for the finest partition of the system into $N$ parties the lifetime $\kappa \tau_N$ essentially scales like $1/N$, for the coarsest partition of the $N$ parties into two groups we have no scaling behavior with $N$, i.e. $\kappa \tau_2 \approx {\rm const}$. 
It follows that in the limit $N\to\infty$ any partitioning in groups of finite size $m$ leads to a vanishing lifetime of the corresponding $M=N/m$ party entanglement. Only if one considers a fixed number of groups $M$ whose size $m=N/M$ grows to infinity as $N\to\infty$, i.e. the groups itself are of macroscopic size, one obtains a finite lifetime of the corresponding $M$--party entanglement.  

We remark that Eq. (\ref{condM}) also allows one to determine the lifetime of genuine $N$--party entanglement, obtained from $M=N/m$ by setting $m=1$.
In a similar way, one can obtain a lower bound on the lifetime of $M$--party entanglement by choosing $\tilde\lambda'_m \equiv 2(1-p)^m(1+p)^{N-m}/2^{N+1} > \lambda_m$. One finds that for $M \leq \log[2(1-p)/(1+p)]/\log[2p/(1+p)]$ all partial transposition with respect to this $M$--partitioning are certainly non--positive, which for these kind of states already ensures that the state is $M$--party distillable \cite{Du00}.
The condition for 2--party entanglement has recently been derived by Simon and Kempe in Ref. \cite{Si02}, who observed that the threshold value for $p$ such that the partial transposition with respect to a partition $N/2$-$N/2$ parties is positive, decreases with the size of the system. Based on this observation, they conclude that GHZ states of more particles are more stable against local decoherence. However, as pointed out in this paper, the effective number of subsystems that remain entangled decreases with time, such that the entanglement becomes bipartite when approaching the threshold value found by Simon and Kempe. The lifetime of genuine $N$--party entanglement thus decreases in fact with the size of the system.

On the other hand, for cluster states (and similar graph states) one can show that there is no scaling with respect to either the size of the partitions or $N$. In this sense both the kind of entanglement in the system and its lifetime are universal. To this aim, it is sufficient to show that there exists not only a lower bound but also an upper bound on the lifetime of entanglement which is independent of $N$ and valid for any partition. To be specific, we show that for any cluster state there exist times $t_{>},t_{<}$ independent of $N$ such that for $t \geq t_{>}$ the state is separable with respect to the finest partition (and hence not distillable with respect to any partition), while for $t < t_{<}$ the state is distillable with respect to the finest partition (and hence distillable with respect to any partition). The existence of such a time $t_{<}$ was already shown earlier in this paper, so we concentrate on $t_{>}$. We use that (i) the graph state $|\Psi_{\vec 0}\rangle$ corresponding to a graph $G$ can be written as \cite{Ra01}
$|\Psi\rangle =\prod_{(k,l)\in E}U_{kl}|+\rangle^{\otimes N}$, where $U_{kl}\equiv e^{-i\pi (\eins^{(k)}+\sigma_z^{(k)})/2\otimes(\eins^{(l)}-\sigma_z^{(l)})/2}$ and $|+\rangle=1/\sqrt{2}(|0\rangle+|1\rangle)$; (ii) The map ${\cal E}_k(p)$ (Eq. (\ref{whitenoise}) with noise parameter $p\equiv e^{-\kappa t}$) can equivalently be described by a concatenation of three maps ${\cal E}_k^{(x)}(p_x) {\cal E}_k^{(y)}(p_y) {\cal E}_k^{(z)} (p_z)$ where ${\cal E}_k^{(j)} (p_j)\rho = p_j\rho+(1-p_j)/2[\rho+\sigma_j^{(k)}\rho\sigma_j^{(k)}]$ with $p_j=\sqrt{p}$, $j=x,y,z$. 
We consider the map $\tilde {\cal E}_{kl}^{(z)}(p_z) \rho \equiv {\cal E}_k^{(z)}(p_z) {\cal E}_l^{(z)}(p_z) U_{kl}\rho U_{kl}^\dagger$. Using the results of Ref. \cite{Ci00}, one can verify that $\tilde {\cal E}_{kl}^{(z)}(p_z)$ is separable and hence not able to create entanglement if $p_z \leq (\sqrt{2}-1)$. Consider now a graph state where each vertex in the corresponding graph has $m$ neighbors. Since $U_{kl}$ commutes with ${\cal E}_j^{(z)}(p_z)$ and ${\cal E}_k^{(z)} (p_z) \rho = {\cal E}_k^{(z)}(\sqrt{p_z}) {\cal E}_k^{(z)} (\sqrt{p_z}) \rho$, we can rewrite $\rho(t)= \prod_{j=1}^N {\cal E}_j^{(x)}(p_x) {\cal E}_j^{(y)}(p_y) \prod_{(k,l)\in E} \tilde {\cal E}_{kl}^{(z)}(p_z^{1/m}) |+\rangle\langle+|^{\otimes N}$. We have that all maps in this expression are separable if $p_z^{1/m} \leq (\sqrt{2}-1)$ which implies that $\rho(t)$ is fully separable for $\kappa t \geq -2m \ln(\sqrt{2}-1)$. This provides the announced upper bound $t_{>}$ on the lifetime of distillable entanglement valid for arbitrary partitions.

In this paper, we have studied the behavior of multiparticle entangled states under decoherence. For GHZ states, we found that the lifetime of true $M$--party entanglement decreases with the size of the system and the effective number of entangled subsystems decreases with time. For cluster and similar graph states, however, we have shown that the lifetime of $N$--party entanglement is independent of the size of the system. 
These results suggests that true multiparticle entanglement in macroscopic objects can be more stable (and might be more common) than previously thought of.

We thank J.I. Cirac for valuable discussions. 
This work was supported by European Union (HPMF-CT-2001-01209 (W.D.), IST-2001-38877,-39227) and the DFG.




\end{document}